\documentclass[
    ,final            
  ]
  {aipproc}

\layoutstyle{8x11double}

\def\lapp{\ifmmode\stackrel{<}{_{\sim}}\else$\stackrel{<}{_{\sim}}$\fi}
\def\gapp{\ifmmode\stackrel{>}{_{\sim}}\else$\stackrel{>}{_{\sim}}$\fi}

\newcommand{\xte}{\textit{RXTE}} 
\newcommand{\tfe}{1E~1048.1--5937} 
\newcommand{\tfn}{1E~2259+586}

\newcommand{\tnin}{\mathrm{T}_{90}}

\begin{document}

\title{A Comprehensive Study of the X-ray Bursts from the Magnetar Candidate \tfn}

\author{Fotis P. Gavriil}{
  address={Department of Physics, Rutherford Physics Building,
McGill University, 3600 University Street, Montreal, Quebec, H3A 2T8,
Canada}
}

\author{Victoria M. Kaspi}{
  address={Department of Physics, Rutherford Physics Building,
McGill University, 3600 University Street, Montreal, Quebec, H3A 2T8,
Canada}
,altaddress={Department of Physics and Center for Space Research,
Massachusetts Institute of Technology,  Cambridge, MA 02139} 
}

\author{Peter M. Woods}{
  address={Space Science Research Center, National Space Science
and Technology Center, Huntsville, AL 35805, USA}
}

\begin{abstract}
We present a statistical analysis of the X-ray bursts observed from the
2002 June 18 outburst of the Anomalous X-ray Pulsar (AXP) \tfn,
observed with the Proportional Counter Array aboard the
\textit{Rossi X-ray Timing Explorer}.  We show that the properties of
these bursts are similar to those of
Soft Gamma-Repeaters (SGRs).  The similarities we find are: the burst
durations follow a log-normal distribution which peaks at 99~ms, the
differential burst fluence distribution is well described by a power law
of index $-1.7$, the burst
fluences are positively correlated with the burst durations, the
distribution of waiting times is well described by a log-normal
distribution of mean 47~s, and the bursts are generally asymmetric with
faster rise than fall times.  However, we find several quantitative
differences between the AXP and SGR bursts.  Specifically, 
there is a correlation of burst phase with pulsed intensity,
the AXP bursts we observed exhibit a wider range of durations,  the
correlation between burst fluence and duration is flatter than for
SGRs, the observed AXP bursts are on average less energetic than observed SGR bursts, and
the more energetic AXP bursts have the hardest spectra -- the opposite of
what is seen for SGRs. We conclude that the bursts are sufficiently
similar that AXPs and SGRs can be considered united as a source class 
yet there are some interesting
differences that may help determine what physically  differentiates the 
two closely related manifestations of neutron stars.
\end{abstract}

\maketitle


\section{Introduction}

Soft gamma repeaters (SGRs) are an exotic class of Galactic sources
that are now commonly accepted as being magnetars -- isolated, young
neutron stars that are powered by the decay of an ultra-high magnetic
field.  The evidence for high surface fields ($\sim 10^{14} -
10^{15}$~G) comes from several independent lines of reasoning
\citep{dt92a,pac92,td95,td96a}.  These include: the high dipolar
magnetic fields implied by the spin properties of SGRs seen in
quiescence  under the assumption of magnetic dipole braking 
\citep{kds+98,ksh+99}; the requirement of a
magnetar-strength field to confine the energy released in the tails of
hyper-Eddington outbursts seen from two SGRs \citep{mgi+79,hcm+99};
the requirement of a high field to allow the decay rate necessary to
power the burst and persistent emission \citep{td96a, gr92}; 
and the magnetic suppression of the
Thomson cross-section, which allows hyper-Eddington bursts to be
observed \citep{pac92}.
For a review of SGRs, see Kouveliotou~et~al.~[these proceedings].

Anomalous X-ray pulsars (AXPs), another exotic class of Galactic
neutron stars, have also been suggested to be magnetars \citep{td96a}.
This is because of their anomalously bright X-ray emission which can
be explained neither by conventional binary accretion models nor
rotation power \citep{ms95}. Also, their spin parameters, as for
SGRs, imply large magnetic fields under standard assumptions of
magnetic braking.  They also have similar, though on average softer,
X-ray spectra compared with those of SGRs in quiescence. However, unlike SGRs, 
in the $>20$~yr since the
discovery of the first AXP \citep{fg81}, none was seen to exhibit SGR-like
bursts.  For this reason, alternative models involving unconventional
accretion scenarios have been proposed to explain AXP emission
\citep{vtv95,chn00,alp01}. See Kaspi~et~al.~[these proceedings] for a review of AXPs.

\begin{figure}
  \includegraphics[width=.36\textwidth]{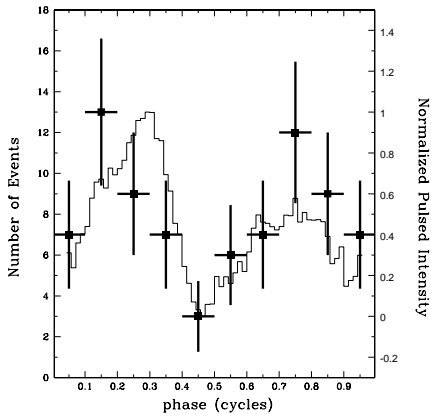}
  \caption{Distribution of the pulse phases of \tfn\ which correspond to the times of the burst peaks (solid points). The solid curve is the folded 2--60~keV light curve of the 2002 June 18 observation with the bursts omitted.
\label{fig:burst phases}}
\end{figure}

The magnetar model for AXPs was recently given a boost
when SGR-like bursts were detected from two AXPs.  \citet{gkw02}
reported on the discovery of two X-ray bursts in observations
obtained in the direction of AXP \tfe.  The temporal and spectral
properties of those bursts were similar only to those seen only in SGRs.
However, the AXP could not be definitely identified as the burster.
On 2002 June 18, a major outburst was detected unambiguously from AXP \tfn,
involving over 80 bursts as well as significant spectral and timing
changes in the persistent emission \citep{kgw+03}.  Those bursts
demonstrated that AXPs are capable of exhibiting behavior observed,
until now, uniquely in SGRs, therefore implying a clear connection
between the two source classes.  Such a connection was predicted only
by the magnetar model \citep{td96a}.

\section{Results}
The results presented here were obtained using the Proportional Counter
Array \citep[PCA;][]{jsg+96} on board the \textit{Rossi X-ray Timing Explorer}
(\xte ).  We use 
\xte\ to monitor all five known AXPs on a regular basis as part of a
long-term monitoring campaign \citep[see][and references therein]{gk02}. 
On 2002 June 18, during one of our
regular monitoring observations, the AXP \tfn\ exhibited an SGR-like outburst
\citep{kgw+03}. 
The bursting behavior was detected by online \xte\ monitors during
the observation.
The observation spanned three orbits and had 
total on-source integration time 10.7~ks. 
Following the outburst, Target of Opportunity observations of the source 
were initiated the next day and continued at different intervals over the subsequent weeks, however no more bursts were seen.

Our burst searching algorithm returned 80 significant bursts 
from the 2002 June 18 observation. Most bursts 
were single-peaked and had durations
$\lapp$1~s.  A small handful ($\sim$12) were bright and had clear
fast-rise, exponential decay morphology.  Four
bursts were  multi-peaked. Some
bursts ($\sim$5$\%$) were approximately symmetric, 
a few ($\sim$3$\%$) fell faster than they rose
while most  fell slower than they rose.

We calculated the occurance in 
pulse phase for each burst using the time of the burst peak and the rotational
  ephemeris given by \citet{kgw+03}.
Comparing  the burst phase distribution to the pulse 
profile of \tfn\  at the time of the outburst, a correlation 
is seen (Fig.~\ref{fig:burst phases}), where most of the  bursts 
tend to occur when the pulsed intensity is high. We note that the two  bursts seen from the AXP \tfe\ \citep{gkw02} were also coincident with the pulse peak, which  strengthens the argument that \tfe\  was the source of those bursts.

\begin{figure}
  \includegraphics[width=.36\textwidth]{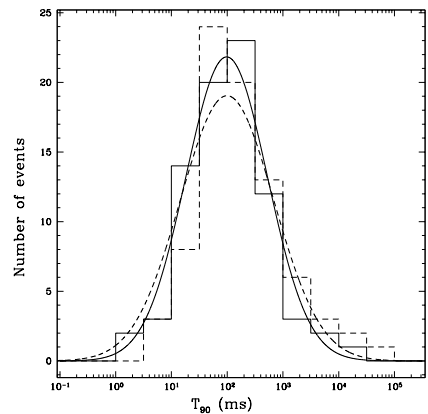}
  \caption{Distribution of $\tnin$ durations  for the bursts observed 
from \tfn.  The solid histogram line shows the observed binned distribution, 
while the dashed histogram line shows the corrected 
distribution (see text).  The solid curve 
represents the best-fit log-normal model for the observed data, while 
the dashed curve is the best-fit log-normal model for the corrected data.  
\label{fig:T90}}
\end{figure}

The $\tnin$ duration is the time
between when $5\%$ and $95\%$ of the total background-subtracted burst
counts have been accumulated \citep[e.g.][]{gkw+01}.  
SGR $\tnin$ distributions follow a log-normal distribution 
whose mean and standard deviation    vary with source \citep[e.g.][]{gkw+01}. 
At first we fit the measured values of $\tnin$ for the
\tfn\ bursts with this model and found it to characterize the
distribution well (Fig.~\ref{fig:T90}). 
For low signal-to-noise bursts, $\tnin$ can be
substantially underestimated.  We corrected for this problem
via Monte Carlo simulations \citep[see][]{gkw03}. The corrected $\tnin$ distribution is shown in Figure~\ref{fig:T90}. 
The best-fit mean is 99.31~ms with  a range of 14.4--683.9~ms
for one standard deviation.

\begin{figure}
  \includegraphics[width=.36\textwidth]{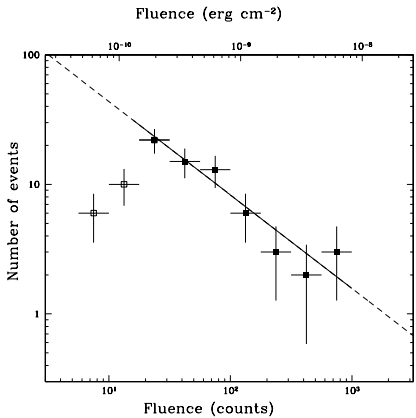}
  \caption{Distribution of
the 2--60~keV fluence  for each burst.  
Solid points represent average values of fluence in equispaced logarithmic
bins for which our observations had full sensitivity.  The open points
suffered from reduced sensitivity.  The best-fit line was determined using
the solid points only and is shown as a solid line; the dashed lines are
its extrapolation.  
\label{fig:fluence distribution}}
\end{figure}

We measured the fluence of each burst and then grouped them  
in equispaced logarithmic bins.  The distribution of burst
fluences is displayed in Figure~\ref{fig:fluence distribution}. The
low-end fluences are underrepresented because of sensitivity drop-off.
Excluding the points having fluence \lapp 20 PCA counts, the
distribution is well modeled by a simple power law. Using least-squares
fitting we find a best-fit power-law index of $-0.7 \pm 0.1$, which 
corresponds to a differential spectrum 
$\mathrm{d}N/\mathrm{d}F \propto F^{-1.7 \pm 0.1}$.
The fluences in the 2--60~keV band 
range from $\sim 5 \times 10^{-11}$ to
$\sim 7 \times 10^{-9}$~erg~cm$^{-2}$. 
These imply burst energies in the range $\sim 5 \times 10^{34}$ to 
$\sim 7 \times 10^{36}$~erg, assuming isotropic emission and 
a distance of 3~kpc to the source \citep{kuy02}.
The sum total of all burst fluences is $5.6\times10^{-8}$~erg~cm$^{-2}$, 
corresponding to 
energy $6.0\times 10^{37}$~erg~(2--60~keV).

\citet{gkw+01} also find a clear correlation between burst durations
and total burst fluence. A similar correlation is seen in our \tfn\ data.  
To quantify it, we grouped the $\tnin$ values in equispaced logarithmic
bins and determined group-averaged fluences for each bin.
Least-squares fitting to a simple power-law model yields $F \propto
\tnin^{+0.54 \pm 0.08}$, with reduced $\chi^2 = 1.0$.

We measured the peak flux in a 61.25-ms time bin for each burst.
Our burst-identifying algorithm is less
sensitive to bursts of smaller peak flux, we compensated for this
effect via simulations \citep[see][]{gkw03}.
Using least-squares fitting we found that the corrected
distribution is well modelled by a simple power law with index $-1.42
\pm 0.13$.
 Peak fluxes in a 61.25-ms time bin  range
from $\sim 1 \times 10^{-9}$ to
$\sim 1 \times 10^{-7}$~erg~cm$^{-2}$~s$^{-1}$, which imply peak luminosities  
in the range $\sim 1 \times 10^{36}$ to $\sim 1 \times 10^{38}$~erg~s$^{-1}$.
On shorter time scales we find 5 bursts with peak fluxes 
which are super-Eddington.  The peak fluxes in a 1/2048~s time bin for these bursts range from $\sim 2 \times 10^{38}$ to $\sim 8\times 10^{38} $~erg~s$^{-1}$.

Burst rise and
fall time distributions were well modelled by 
log-normal distributions.  For the rise time distribution, 
we find a mean of $2.43$~ms and a range of 0.51--11.51~ms for 
one standard deviation.  
For the fall time distribution,
we find mean 13.21~ms and a range  of 3.52--49.55~ms for one
standard deviation.

SGR waiting times, defined as the temporal separations of
adjacent bursts, are found to follow log-normal distributions
\citep{gwk+99,gwk+00}.  We measured the waiting time for the
\tfn\ events, excluding those interrupted by Earth occultations.
Our waiting time distribution is also well modelled by 
a log-normal distribution with mean 
46.7~s and a range of 10.5--208.4~s for one
standard deviation. 
We find no correlation between the burst energy, 
duration and the waiting time until the
next burst, nor with the elapsed time since the previous burst.
There is a  correlation between the waiting time and the event time, 
which  implies that 
the mean of  our waiting time distribution depends on the time at which 
we started  observing the outburst.

\begin{figure}
  \includegraphics[width=.36\textwidth]{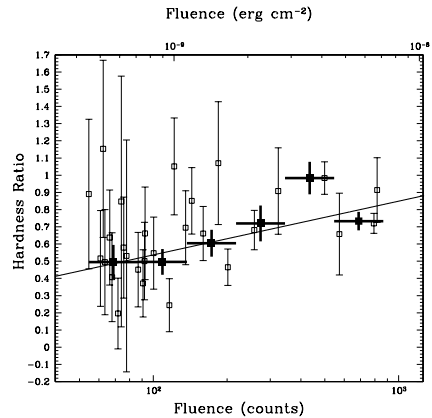}
  \caption{Hardness ratio ($H$) versus fluence ($F$).  
The open points are hardness ratio measurements
for individual bursts. The solid points are weighted averages of
hardness ratios for bursts in equispaced logarithmic fluence bins. The line represents the best-fit logarithmic function for the weighted averages, $H = 0.31\times \log F - 0.09 $.
\label{fig:hardness ratio versus fluence}}
\end{figure}

\citet{gkw+01} noted that SGR bursts tend to soften with increasing
burst energy.  We studied the hardness ratio/fluence relationship by
extracting spectra and creating response matrices separately for each
burst.  Hardness ratios were defined as the ratio of the counts in the
10--60~keV band to those in the 2--10~keV band as in \citet{gkw+01}.
Also following \citet{gkw+01}, we divided the bursts into equispaced
logarithmic fluence bins and calculated a weighted average hardness
ratio for each bin. Figure~\ref{fig:hardness ratio versus fluence}
shows the weighted mean hardness ratios as a function of fluence.  A
clear positive correlation is seen.

\section{Discussion}

The bursts we have observed for \tfn\ are clearly similar to those seen 
uniquely in SGRs.  As concluded by \citet{gkw02} and \citet{kgw+03},
AXPs and SGRs clearly share a common nature, as has been predicted by
the magnetar model.
The bursts seen in the
2002 June 18 outburst of \tfn\ are qualitatively similar to
those seen in SGRs, and in many ways quantitatively similar.
Specifically:
\begin{itemize}
\item the mean burst durations are similar
\item the differential burst fluence spectrum is well described 
by a power law of index $-1.7$, similar to those seen in SGRs (and earthquakes
and solar flares)
\item burst fluences are positively correlated with burst durations
\item the distribution of and mean waiting times are similar
\item the burst morphologies are generally asymmetric, with rise times usually
shorter than burst durations
\end{itemize}
However, there are some interesting quantitative differences between
the properties of the AXP and SGR bursts.  These may help shed light on
the physical difference(s) between these classes.  The differences can be summarized as:
\begin{itemize}
\item there is a significant correlation of burst phase with pulsed intensity, unlike in SGRs 
\item the AXP bursts have a wider range of burst duration (though this may be partly due to different analyses procedures)
\item the correlation of burst fluence with duration is flatter for AXPs than
it is for SGRs (although when selection effects are considered, this correlation
should really be seen as an upper envelope for AXPs and SGRs)
\item the fluences for the AXP bursts are generally smaller than are in observed SGR bursts
\item the more energetic AXP bursts have the hardest spectra, whereas for SGR bursts,
they have the softest spectra
\item under reasonable assumptions, SGRs undergo outbursts much 
more frequently
than do AXPs
\end{itemize}

Given the rarity of AXP bursts coupled with the unique
information that detection of such bursts provides, observing more
outbursts is obviously desirable.  Continued monitoring is thus clearly
warranted, and \xte\ with its large area and flexible scheduling
is the obvious instrument of choice.


\begin{theacknowledgments}
We are grateful to C.~Kouveliotou, M.~Lyutikov, S.~Ransom,
M.S.E.~Roberts, D.~Smith, and C.~Thompson for useful discussions.
This work was supported in part by NSERC, NATEQ, CIAR and NASA.
This research has
made use of data obtained through the High Energy Astrophysics Science
Archive Research Center Online Service, provided by the NASA/Goddard
Space Flight Center.
\end{theacknowledgments}


\bibliographystyle{aipproc}   



\IfFileExists{\jobname.bbl}{}
 {\typeout{}
  \typeout{******************************************}
  \typeout{** Please run "bibtex \jobname" to optain}
  \typeout{** the bibliography and then re-run LaTeX}
  \typeout{** twice to fix the references!}
  \typeout{******************************************}
  \typeout{}
 }

\end{document}